\documentclass{aa}  

\usepackage{graphicx}
\usepackage{txfonts}
\usepackage{lipsum}
\usepackage{subcaption}         
\usepackage{ulem}         
\usepackage{lscape}
\usepackage{placeins}
\usepackage{xcolor}

\newcommand\halpha{H$\alpha$}
\newcommand\caii{Ca~{\sc ii}}
\newcommand\siiv{Si~{\sc iv}}
\newcommand\mgiik{Mg~{\sc ii} k}

\newcommand\kmps{km\,s$^{-1}$}
\newcommand\kmpss{km\,s$^{-2}$}

\newcommand\tv{$t$-$v_D$}
\newcommand\vs{$v_D$-$s$}
\newcommand\ts{$t$-$s$}

\begin{document}

\title{Spectroscopic observations of periodic bursty ejections from a dome-spire-shaped jet}

   \author{Heesu Yang\inst{1}\fnmsep\thanks{Corresponding author: hsyang@kasi.re.kr}
        \and Maria S. Madjarska\inst{2,1,3}
        \and Kyoung-Sun Lee\inst{4,5}
        \and {Klaus Galsgaard}\inst{6}
        }

   \institute{Korea Astronomy and Space Science Institute, 776 Daedeok-daero, Yuseong-gu, 34055, Daejeon, Republic of Korea
   \and Max Planck Institute for Solar System Research, Justus-von-Liebig-Weg 3, 37077, G\"{o}ttingen, Germany
   \and Space Research and Technology Institute, Bulgarian Academy of Sciences, Acad. G. Bonchev Str., Bl. 1, 1113, Sofia, Bulgaria
   \and Astronomy Program, Department of Physics and Astronomy, Seoul National University, Gwanak-gu, 08826, Seoul,
Republic of Korea
   \and Astronomy Research Center, Seoul National University, Gwanak-gu, 08826, Seoul, Republic of Korea
   \and School of Mathematics and Statistics, University of St Andrews, North Haugh, St Andrews, KY16 9SS, Scotland, UK}
   
   \date{}

  \abstract  
   {The study investigates the complexity of the formation and evolution of a dome-spire-shaped jet (or anemone jet) that originated in an active region and was presumably driven by null-point magnetic reconnection.}
  {We aim to provide a quantitative estimation of multi-temperature plasma dynamics and physical parameters of the jet and its blob-like features, together with its temporal evolution in 3D.}
  {For this, we analyzed unique sit-and-stare spectral (Si\,{\sc iv}, \mgiik, and C~{\sc ii}) and slit-jaw imaging 1400~\AA\ and 2796~\AA\ observations obtained with the Interface Region Imaging Spectrograph (IRIS), with a  slit uniquely positioned just above the assumed reconnection site. Co-observations taken with the Fast Imaging Solar Spectrograph of the Goode Solar Telescope in the chromospheric H$\alpha$ and Ca~{\sc ii}~8542.10\,\AA\ lines were also analyzed. These data are complemented by extreme-ultraviolet imaging information from the Atmospheric Imaging Assembly on board the Solar Dynamics Observatory.}
  {The IRIS observations reveal that the jet is formed by multiple periodic ($\sim$30~s) plasma ejections with blob-like appearances, each associated with strongly blueshifted Si\,{\sc iv} 1402.77\,\AA\ transition-region emission that reaches about $-300$\,\kmps. More importantly, the ejections are grouped into well-defined packets, each lasting approximately 3--5 min. In contrast, only a single Doppler component is observed in the chromospheric \caii\ line. Multi-Gaussian decomposition of the Si\,{\sc iv} line profiles reveals a highly accelerated plasma, with the line-of-sight speed toward the observer increasing linearly from $\sim$30 to 112\,\kmps~over 66\,s, corresponding to an acceleration of 1.3\,\kmpss. The kinetic energy release rate is estimated to be in the range $\sim$1.0$\times10^{24}$--1.0$\times10^{25}$ erg\,s$^{-1}$. }
{}
\titlerunning{Periodic bursty ejections during a dome-spire-shaped jet}

\maketitle
\nolinenumbers

\section{Introduction}
\label{sec:intro}

Solar jets are dynamic collimated plasma ejections that can originate in the chromosphere, transition region, or corona. Some exhibit fan–spine structures and are referred to as anemone, inverted-Y, or $\lambda$-shaped jets; they are observed in a wide range of wavelengths and temperatures, including X-ray and extreme-ultraviolet (EUV) observations \citep[e.g.,][]{shibata_1992,canfield_1996,jiang_2007}. From a 3D perspective, these jets are interpreted as originating from a dome-spire–shaped magnetic configuration and are associated with magnetic reconnection occurring at a null point located above the parasitic polarity region enclosed by the dome \citep[e.g.,][]{pontin_2013}.

The dome–spire structure, composed of spine and fan field lines, has been extensively investigated as a model for 3D magnetic reconnection through magnetohydrodynamic simulations and analytical solutions \citep[e.g.,][]{pontin_2013,pontin_2022}, as it represents one of the simplest forms of 3D magnetic topology to host a null point. In this structure, a coronal null point is located above the dome, where the magnetic field locally vanishes and stress accumulates in response to external perturbations, such as photospheric shear. As a result, a current sheet forms around the null point; this triggers spine-fan reconnection, which transfers magnetic flux across the separatrix dome and drives plasma ejections along the spire \citep{pontin_2022}.

High-resolution observations strongly support this reconnection scenario of jet formation. Brightenings along the fan surface have been interpreted as signatures of current-sheet formation at the separatrix or quasi-separatrix layers. The observation of recurrent bi-directional flows across the separatrix surface and along the spine has also been suggested as evidence of magnetic reconnection at the separatrix layers or null points \citep[e.g.,][]{zhang_2014,zeng_2016,kumar_2019}. 

Magnetic reconnection in the Sun is not steady but inherently impulsive and bursty. This is a natural consequence in resistive regimes, arising from time-dependent reconnection \citep[for review, see][]{priest_2007}. The current sheet formed at the null point is subject to plasmoid instability---a rapidly growing secondary tearing of a Sweet–Parker current sheet---which generates multiple plasmoids and secondary current sheets \citep{bhattacharjee_2009}. In dome–spire jets, the stretching of the null region forms a vertical current sheet \citep{edmondson_2010}, which can become unstable due to the tearing instability \citep{kliem_2000, ni_2015}; this lead to bursty, quasi-periodic outflows that are frequently observed in solar jets \citep{mulay_2023}.

Interestingly, recent high-resolution Interface Region Imaging Spectrograph \citep[IRIS;][]{depontieu_2014} and Solar Dynamics Observatory \citep[SDO;][]{Pesnel_2012} observations have confirmed the presence of elongated blob-like features moving along the spire of jets. Their apparent widths range from 0.3 to 2.3~Mm \citep{kumar_2019,zhang_2019,chen_2022,mulay_2023}, and their velocities from 50 to over 300\,\kmps~during impulsive phases \citep{kumar_2019,joshi_2020,chen_2022}. These blob-like features often exhibit multi-temperature properties (1--2.2 MK; \citealt{mulay_2023}) and are sometimes cooler than the surrounding plasma \citep{kumar_2019}. They have been interpreted as plasmoids; however, their physical nature remains poorly constrained, mainly because it is difficult to disentangle their dynamics from overlapping large-scale flares or unresolved substructures.

In this study, we investigated a collimated cool and hot plasma flow (a jet) emanating from a dome-shaped base observed in spectroscopic and imaging data taken by IRIS and the Fast Imaging Solar Spectrograph \citep[FISS;][]{chae_2013} installed at the Goode Solar Telescope (GST) of Big Bear Solar Observatory \citep[][]{wenda_2010}. Thanks to combined chromospheric and transition-region spectroscopic and imaging data, we were able to investigate recurrent blob-like plasma ejections, jet morphological properties, Doppler velocities, proper-motion speeds, jet acceleration, cool-jet component dynamics and properties, and the kinetic energy release rate in detail.

\section{Observation and data analysis\label{sec:obs}}

The observations were carried out on 2024 August 6 from 16:29~UT to 21:29~UT with IRIS, while the FISS data cover the time period from 18:40~UT to 21:13~UT. The coordinates of the targeted jet were (–597\,\arcsec, –262\,\arcsec) in helioprojective coordinates at 21:04:11~UT, corresponding to $\mu \approx 0.71$ ($\theta\approx 44^\circ$).

IRIS provides high-resolution slit-jaw images (SJIs) in the 1400\,\AA\ and 2796\,\AA\ channels, covering a raster field of 120\arcsec$\times$119\arcsec~with a cadence of 19\,s. The EUV and ultraviolet (UV) spectra were obtained in a sit-and-stare mode with a cadence of 9.4\,s, a spectral resolution of $\sim$26 m\AA, and spatial sampling of 0.33\arcsec\,pixel$^{-1}$. FISS simultaneously observed the chromospheric lines \halpha~6562.83\,\AA~and \caii~8542.10\,\AA, scanning a 32\arcsec$\times$39\arcsec~field of view with a cadence of about $30$\,s and spatial sampling of 0.16\arcsec. The FISS data reduction followed the pipeline described in \citet{chae_2013}. We also used data from the Atmospheric Imaging Assembly \citep[AIA;][]{lemen_2012} on board SDO. The AIA EUV data have a cadence of 12~s and a pixel size of 0.6\arcsec$\times$0.6\arcsec.

We performed sub-pixel co-alignment between AIA 1700\,\AA~images and far wing raster images of the FISS \halpha~line using cross-correlation techniques. We compared the photospheric patterns observed in the AIA 1700\,\AA~images with those in the far-wing \halpha~raster images. Lastly, the overall co-alignment was visually inspected.

\section{Results}

\subsection{Jet's morphological properties and velocity}\label{morphology}
\begin{figure}[ht!]
\includegraphics[width=\hsize]{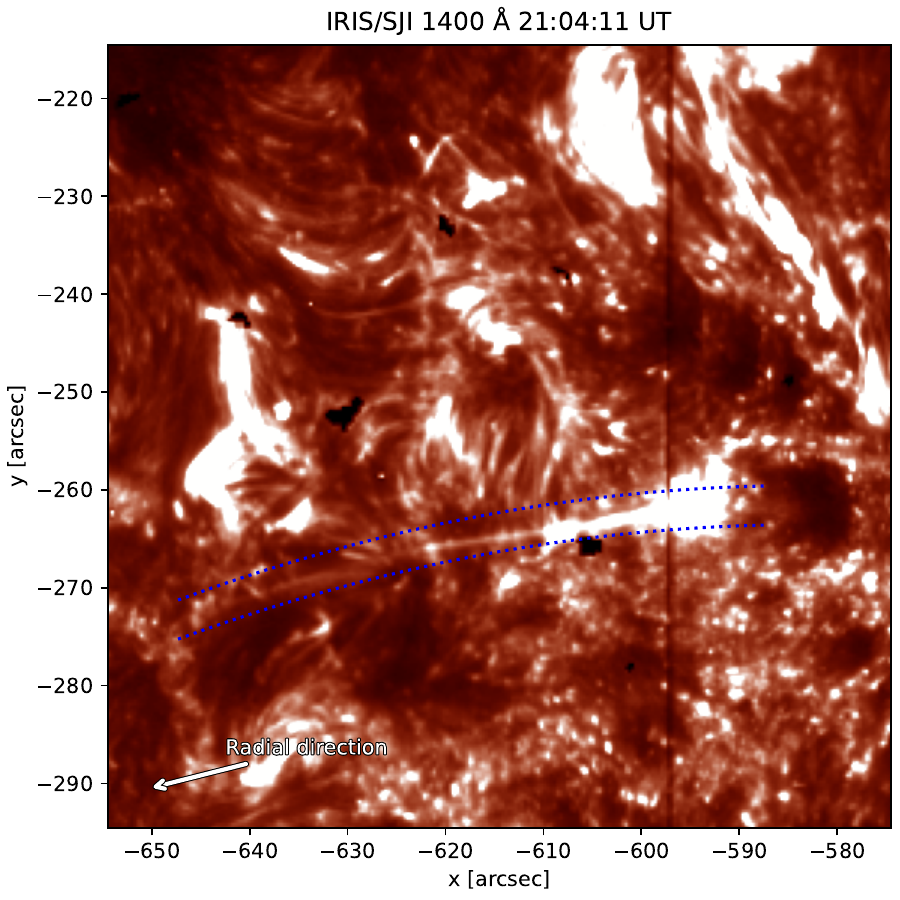}
\caption{IRIS SJI 1400\,\AA\ image taken at 21:04:48~UT, showing the observed anemone jet (dotted~blue lines). The corresponding animation of the complete observing sequence is available online.
\label{fig:fov}}
\end{figure}

The jet originates in the active region NOAA 13780 and is located near pores and is associated with a parasitic polarity embedded in a unipolar region of opposite sign (see Fig~\ref{fig:mag}).
Figure~\ref{fig:fov} and the accompanying animation present the spatial evolution of the jet from its onset. The jet first appeared at around 19:45 UT and lasted longer than 100~min, i.e., until the end of the observations at 21:28 UT. 
The jet presents as a bright, highly collimated flow along a curved trajectory with a length exceeding 50\arcsec\ in the slit jaw (SJ) 1400\,\AA~images (see the dotted blue lines in Fig.~\ref{fig:fov}).

\begin{figure*}[ht!]
\centering
\includegraphics[width=\hsize]{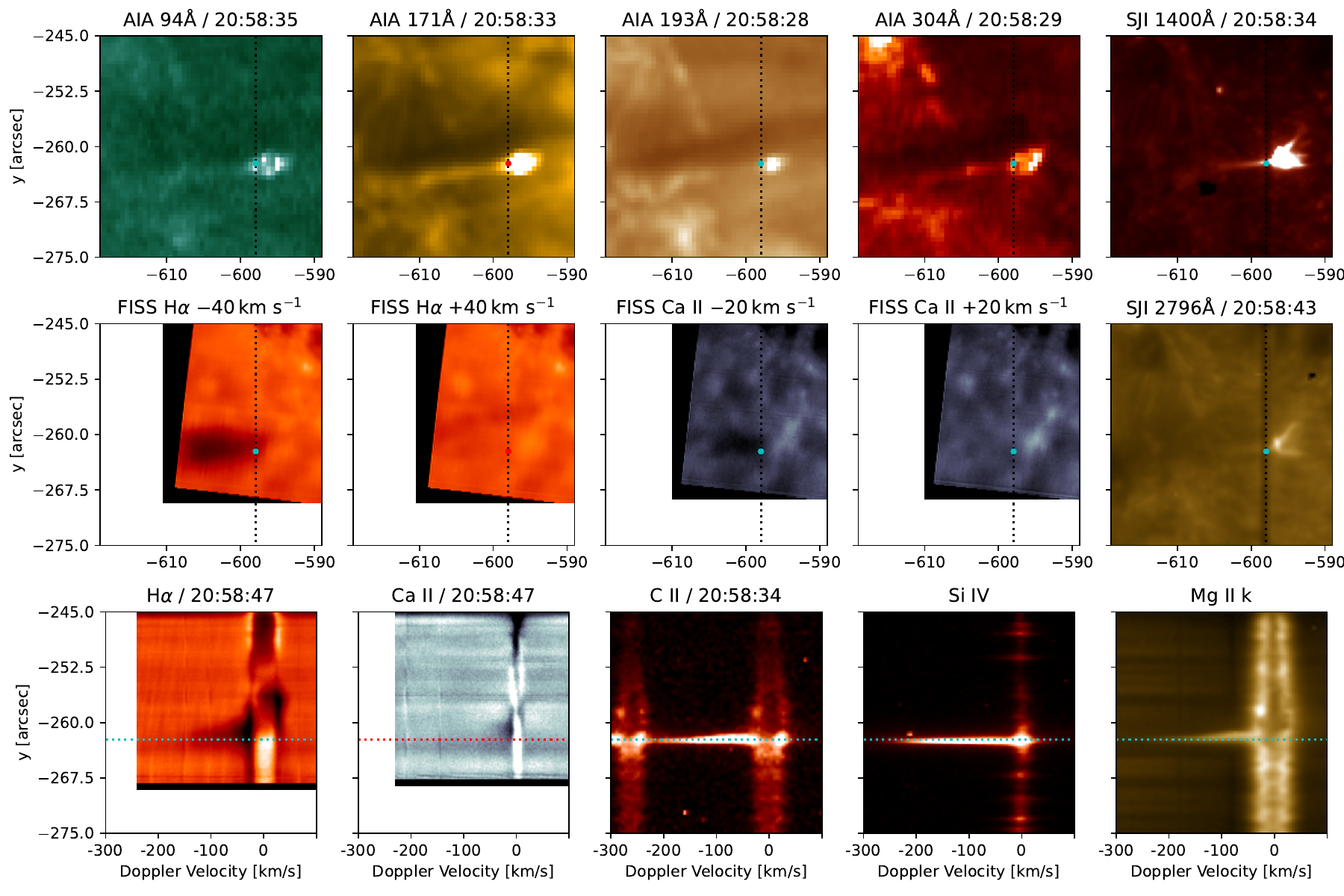}
\caption{{\textit{Top} and \textit{middle panels}:} Multiwavelength view of the observed jet. The vertical dotted lines indicate the IRIS slit position, which is shown in the bottom panels. {\textit{Bottom panels}:} Corresponding Doppler velocity--slit height (\vs) images within the field of view. The horizontal dotted  lines indicate the locations where the spectra used in the time--Doppler velocity (\tv) plots in Fig.~\ref{fig:wtplot} are extracted. An associated animation is available online.
\label{fig:multiobs}}
\end{figure*}

Figure~\ref{fig:multiobs} and the accompanying animation present a multiwavelength view of the observed jet together with the slit spectra. The slit is located at the tip of the dome, i.e., just above the assumed null point. 
Prior to 20:04~UT, a cool plasma jet is seen in absorption in AIA~171, 193, 211, 304\,\AA, \halpha, and \caii~blue wing images. 
In the \halpha\ and \caii\ images, a bright feature located in the footpoint of the jet (no shape can be distinguished due to resolution limitations) is observed at around 20:08~UT (see the animation).  The cool jet, seen in absorption across all AIA EUV channels, persists throughout the observations. The jet also manifests as a redshifted absorption located in the northern part of the jet seen later in the SJI 1400~\AA\ and in the \halpha~Doppler velocity--slit height (\vs)~images in Fig.~\ref{fig:multiobs}. We interpret the redshifted absorption as predominantly downflowing cool plasma associated with the jet. This is also evident in the C~{\sc ii} and \mgiik~spectra, where it is identified as a redshift of the line core. After 20:30~UT, the \halpha\ and \caii\ bright feature disappears, and only the collimated dark absorption feature is visible in the \halpha~and \caii~blue wing images. During the observations, the seeing conditions were not stable, and the jet was located away from the aplanatic patch of the adaptive optics, resulting in a hazy appearance in the \halpha~and \caii~images.

\begin{figure*}[ht!]
\includegraphics[width=\hsize]{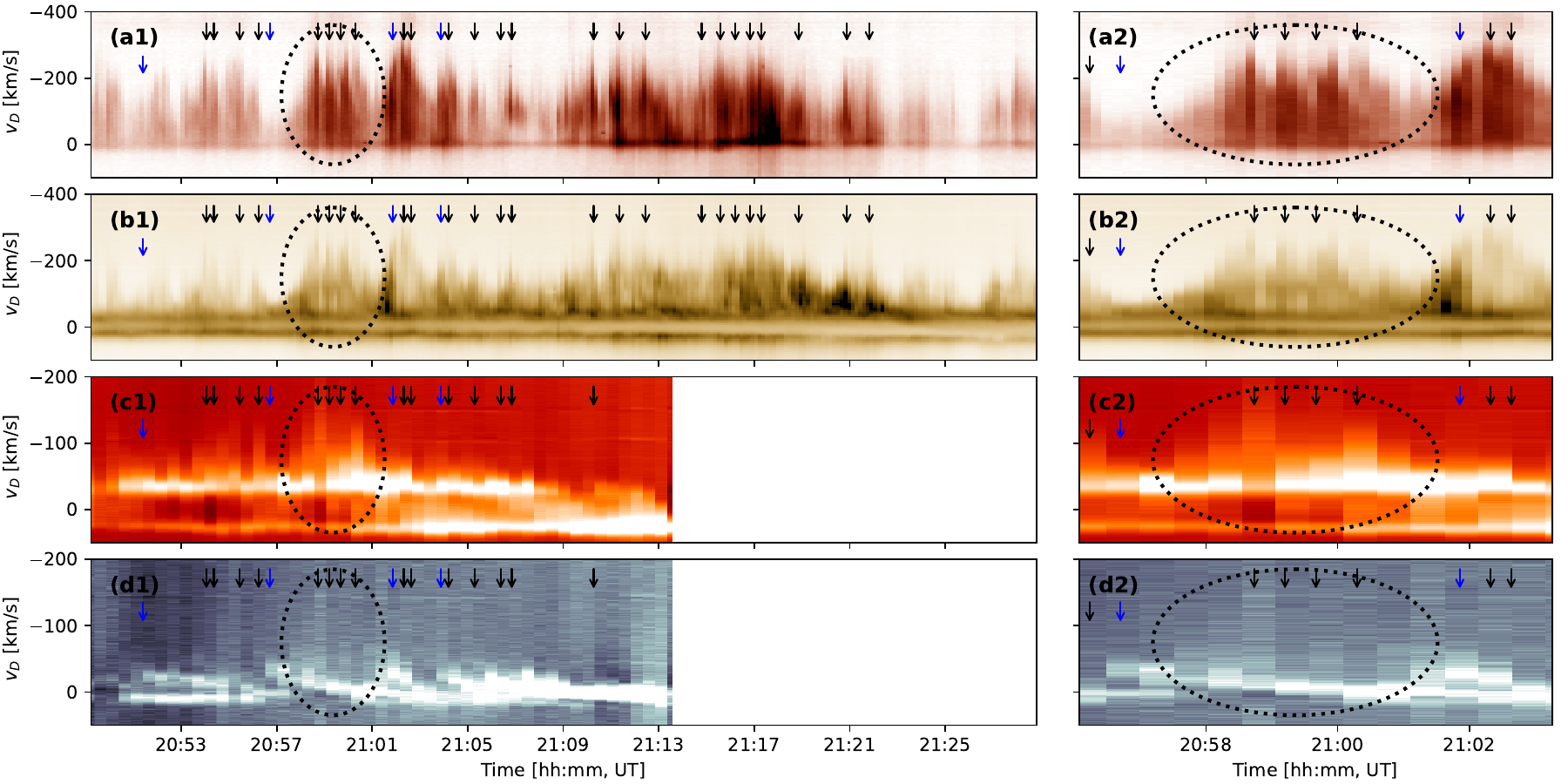}
\caption{Time--Doppler velocity (\tv) plots of the Si\,{\sc iv}~1402.77\,\AA\ (a1), Mg\,{\sc ii}\ k\ 2796.35\,\AA\ (b1), H$\alpha$\ (c1), and Ca\,{\sc ii}\ (d1) lines extracted at the spire locations marked in Fig.~\ref{fig:multiobs}. The black arrows mark the individual ejections best seen in the Si\,{\sc iv}~1402.77\,\AA\ channel, and the blue arrows point at the sawtooth pattern seen in the \caii\ spectral data. The right panels (a2--d2) show the zoomed-in views of the \tv~plots. Note that reversed color tables are used in all panels.
\label{fig:wtplot}}
\end{figure*}

\begin{figure*}[ht!]
\includegraphics[width=\hsize]{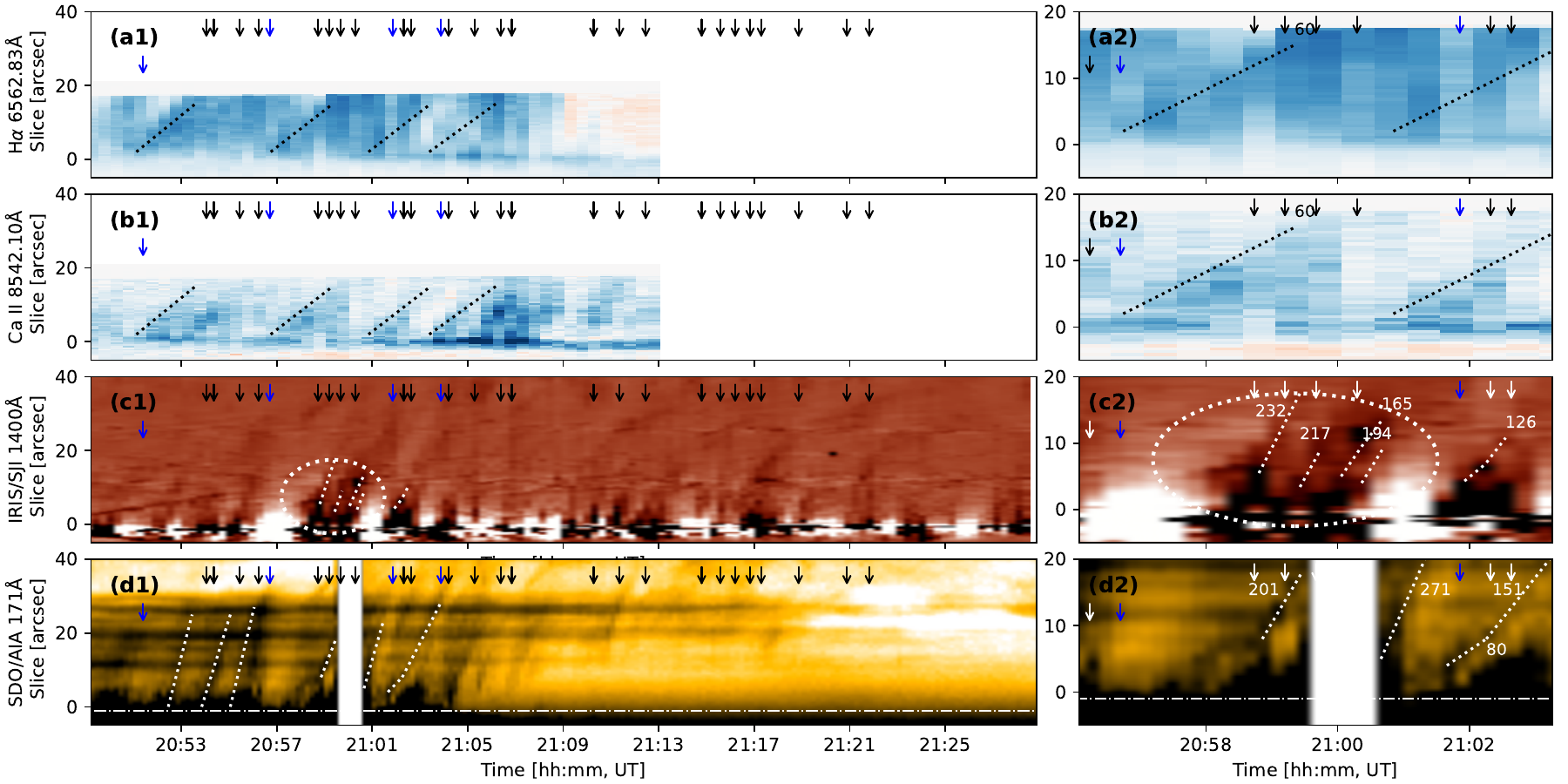}
\caption{Time-slice (\ts) plots along the jet spire marked in Fig.~\ref{fig:fov}, derived from the pseudo velocity of  H$\alpha$$\pm40$\,km\,s$^{-1}$ (panels a1 and a2) and Ca\,{\sc ii} $\pm20$\,km\,s$^{-1}$ (b1 and b2) images, and from the intensity of IRIS/SJI 1400\,\AA~images (c1 and c2)  and AIA 171\,\AA~images (d1 and d2).  Dotted black and white lines denote the trajectories of features propagating along the spire in each dataset. The arrows mark the same interval as in Fig.~\ref{fig:wtplot}. The color tables of the bottom two rows (c1, c2, d1, and d2) are reversed, and panels (d1) and (d2) show the intensity on a logarithmic scale.
\label{fig:stplot}}
\end{figure*}

After 20:04~UT, a bright dotted patch emerges in the SJ 1400\,\AA~image evolving into an inverted-$\lambda$ structure. This structure shows intermittent brightness, as seen in the SJ~1400\,\AA, AIA~171, 193, 211, and 304\,\AA\ channels, expelling hot plasma (appearing bright), i.e., a high-temperature jet. Thus, these observations reveal the formation and evolution of a jet that has both cool and hot plasma components. These jets evolved over a period of 100\,min, starting with a smaller jet that evolves into a very dynamic, bright ejection. In the following, we report an in-depth imaging (Figs.~\ref{fig:multiobs} and \ref{fig:stplot}) and spectroscopic analysis (Figs.~\ref{fig:multiobs}, \ref{fig:wtplot}, \ref{fig:acc_sp}, \ref{fig:spfit}, and \ref{fig:fitres}) of the most dynamic period of the jet evolution.

Around 20:58~UT, the SJ~1400\,\AA\ and 2796\,\AA\ images clearly show a dome-shaped footpoint structure, with only a few pixels from this dome structure showing increased brightness (see Fig.~\ref{fig:multiobs} and the accompanying animation). The jet reached its peak activity around 21:00~UT after gradually becoming more active. The hot jet is only visible in the AIA EUV channels until 21:05~UT, as the evolving cool jet moves in front of it along the line of sight, obscuring the high-temperature emission. The bright dome-shaped jet structure remains visible in the SJ~1400\,\AA. The SJ 2796\,\AA\ images also show a faint, dome-shaped, emissive feature with a spire, along with a slightly dimmer feature on the northern side that resembles the dark structure seen in the AIA EUV channels.

The width of the cool component traced in the \halpha, \caii, and AIA EUV absorption is distributed over a broader area than the collimated emissive hot spire, and the darkest region of the cool component appears offset from the emissive spire. Such a spatial distribution, in which the components are adjacent to but offset from each other, is consistent with observational \citep[][and references therein]{shen_2021} and numerical studies \citep{yokoyama_1996}.

At the start of the jet, the dome-shaped bright patch is located about 3\arcsec\ from the SJ slit. Later, the dome expands, moving closer to the slit location at a distance of no more than 1\arcsec\,(see the dash-dotted  white lines in Figs.~\ref{fig:stplot}d1 and \ref{fig:stplot}d2, which mark the location of the AIA 171\,\AA~brightenings). The width of the narrow spire of the jet identified in the SJ 1400\,\AA~images remained nearly constant during the observations. It has a small curvature toward the north, as indicated by the dotted blue line in Fig.~\ref{fig:fov}, and its shape and position were almost unchanged throughout the observations. The jet width is estimated at $w_{\rm{FWHM}}\sim0.76\arcsec\pm0.25\arcsec$, which is only about twice the IRIS instrumental diffraction limit of 0.33\arcsec. The spire location was determined at sub-pixel resolution at each time step by fitting the intensity profile averaged over the velocity range of 100--130\,\kmps~along the slit.

Figure~\ref{fig:wtplot} displays the variations of the spectra extracted at the location of the jet spire in the Si~{\sc iv}~1402.77\,\AA, \mgiik~2796.35\,\AA, \halpha, and \caii~lines.
The figure clearly shows periodic blueshifted ejections crossing the slit over time. These ejections are grouped into well-defined packets, each lasting approximately 3--5~min, as seen in Figs.~\ref{fig:wtplot}a and \ref{fig:wtplot}b. For instance, a packet appearing around 20:57~UT (marked with a dotted black ellipse) consists of at least four blueshifted ejections (marked with the black arrows), which are separated by short intervals of roughly 30~s (3--4 frames) showing emission decrease.

One notable feature is that each packet corresponds to a sawtooth pattern, described as a blueshifted pattern suddenly appearing, followed by deceleration, seen in the Ca~{\sc ii} time--Doppler velocity (\tv)\ plot in Fig.~\ref{fig:wtplot}d. Such sawtooth patterns are identified at around 20:51, 20:57, 21:02, and 21:04~UT (marked with blue arrows). This velocity pattern resembles that of surges previously seen in imaging-spectroscopic observations \citep[see Fig.~2 of][]{yang_2014}.

The blueshifted spectral pattern in the H$\alpha$~\tv~plot appears similar to that observed in the Si~{\sc iv} and Mg~{\sc ii} plots, but not that of the \caii. Its velocity, however, reaches only a few tens of \kmps, which is less than the blueshifted emission observed in the IRIS spectra. This discrepancy is consistent with previous reports \citep[e.g.,][]{chae_1999,jiang_2007} suggesting hot jets move faster than cool jets. 

 Figure~\ref{fig:stplot} displays the jet's proper motion in the \halpha, \caii, SJ~1400\,\AA, and AIA~171\,\AA\ time-slice (\ts) images extracted from an image slice along the jet-spire marked in Fig.~\ref{fig:fov}. We subtracted the background from panels (c1) and (c2), which we estimated with a 1D Gaussian filter applied along the time axis to enhance the visibility of the moving features (see Appendix~\ref{appx:gaussian_filter}). The filter has a standard deviation $\sigma=5.6$~pixels. In the \halpha\ and \caii\ \ts\ images the jet has a speed of about 60\,\kmps, while in the SJ 1400\,\AA~and AIA 171\,\AA~\ts\ images its speed is of about 120--300\,\kmps. This EUV projection speed range is likely consistent with the previous studies (59--443\,\kmps; \citealt{zhang_2016,kumar_2019,joshi_2020,chen_2022,mulay_2023}). Notably, the jet in SJI~1400\,\AA\ does not appear as a collimated continuous flow, but rather has a discrete and fragmented structure of intermittently expelled blob-like features along the jet spire, as shown in Fig.~\ref{fig:stplot}(c2). At least four bright blob-like features are ejected during the packet that starts at 20:57~UT, which is consistent with the blueshifted ejections identified in Fig.~\ref{fig:wtplot}. The lengths of the blob-like features are around 2--3\arcsec, measured as described in Appendix~\ref{appx:blobsz}.

\subsection{Spectral properties of the jet's blob-like features}

To investigate the properties of the blob-like ejecta, we performed a multi-Gaussian fit to the Si~{\sc iv}~1402.77\,\AA~spectra at each time step under the assumption of optically thin emission. Each spectrum was fitted with one, two, or three Gaussian components (G1, G2, and G3). The number of Gaussian components was determined manually based on the number of peaks identified in the spectra.

A stationary Gaussian component (G1) at the \siiv\ 1402.77\,\AA\ line center is associated with the background emission. We fixed this component centered at 1402.77\,\AA\ ($\lambda_1$) with a Gaussian width ($\sigma_1$) of $52$\,m\AA, corresponding to a nonthermal speed of 14\,\kmps~by adopting the thermal broadening ($\sigma_{\rm{th}}$) of \siiv~formed in the transition region at a temperature of about 80\,000\,K and the instrument broadening ($\sigma_{\rm{inst}}$) of 1.97\,\kmps\,\citep{depontieu_2015}. We determined these fixed values, $\lambda_1$ and $\sigma_1$, as follows: First, we performed an initial fit for each profile without any fixed values. After that, $\lambda_1$ and $\sigma_1$ were obtained by averaging the initial fitting results. Therefore, in the fitting, the only free parameter of G1 is its amplitude, $A_1$. As a result, the line shift, width, and amplitude of G2 and G3 ($A_2$, $\lambda_2$, $\sigma_2$, $A_3$, $\lambda_3$, $\sigma_3$), together with $A_1$, were determined by fitting each spectrum.

\begin{figure}[ht!]
\centering
\includegraphics[width=\hsize]{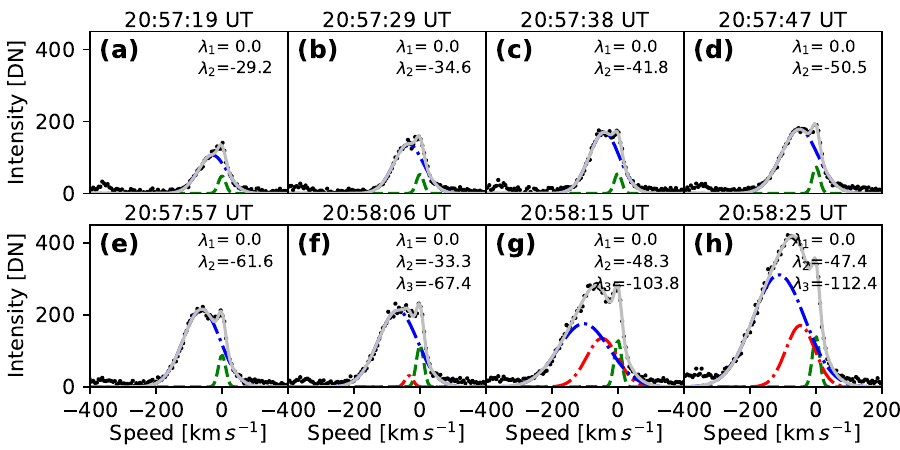}
\caption{Si\,{\sc iv} spectra and the corresponding Gaussian fits at each time step within the acceleration period from 20:57:19~UT to 20:58:25~UT. The black dots mark the observed data, and the solid gray line marks the fitting result. The Gaussian components G1, G2, and G3 are indicated by dashed green, dash-dotted red, and dash-dotted blue lines, respectively. When the fit was performed with two Gaussians, the most blueshifted component was assigned to G3.\label{fig:acc_sp}}
\end{figure}

Figure~\ref{fig:acc_sp} displays the \siiv~spectra and the corresponding Gaussian fit at each time step within the time interval from 20:57:19~UT to 20:58:25~UT. In this example, the spectra from panels (a) to (e) are fitted with two Gaussians, whereas the spectra from (f) to (h) are fitted with three Gaussians. An emission component appears in the blue wing and progressively shifts farther into the wing. This corresponds to an enhancement of the blueshift from approximately 30 to 112~\kmps~over a period of about 66~s, corresponding to an acceleration of $\sim$1.3~km\,s$^{-2}$ (we derived this value from a linear fit to the data points in the first period marked with a double-bar arrow in Fig.~\ref{fig:fitres}a). 
The amplitude and the Doppler width ($A_3$ and $\sigma_3$) also continuously increase during the acceleration period (Figs.~\ref{fig:fitres}b and \ref{fig:fitres}c). 

\begin{figure}[ht!]
\includegraphics[width=\hsize]{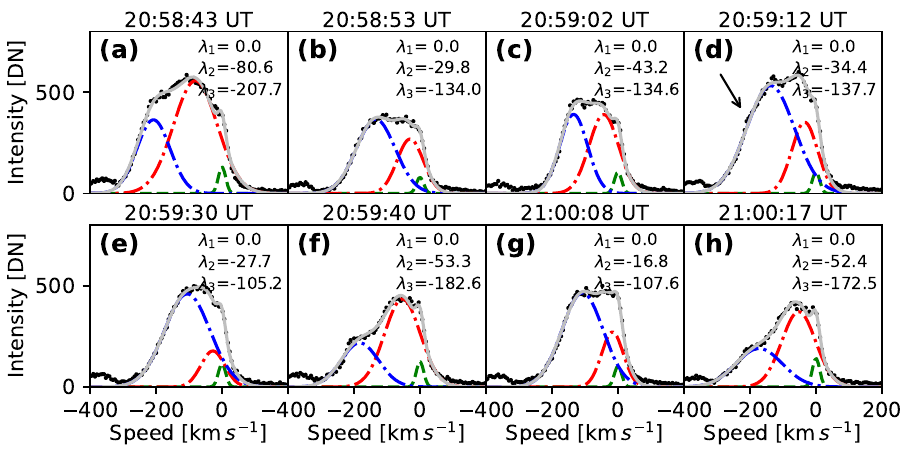}
\caption{Si\,{\sc iv} spectra at each time step immediately following the time step shown in Fig.~\ref{fig:acc_sp}. The line colors and the styles are the same in Fig.\ref{fig:acc_sp}.\label{fig:spfit}}
\end{figure}

\begin{figure}[ht!]
\includegraphics[width=\hsize]{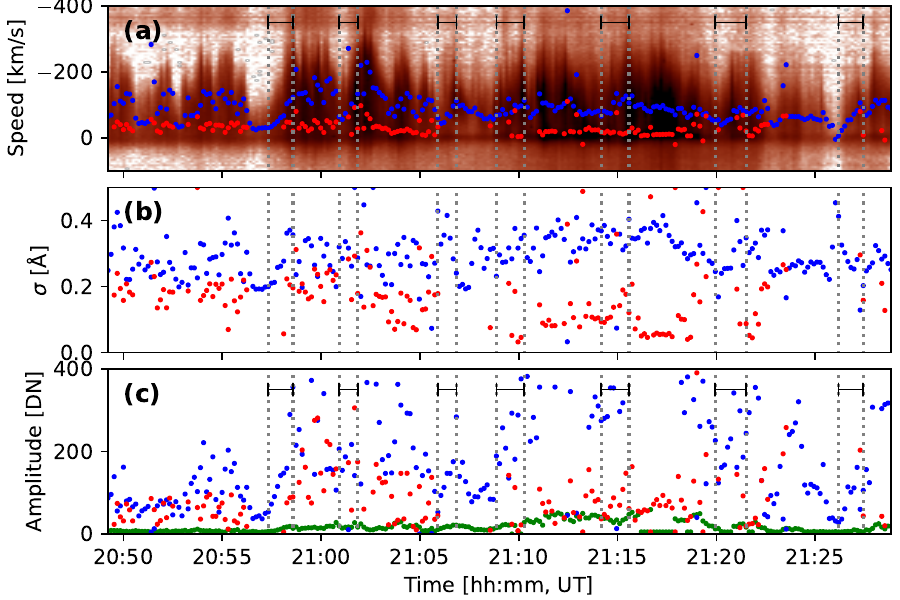}
\caption{Multi-Gaussian fitting results for the Si\,{\sc iv}\ 1402.77\,\AA\ spectra along the jet spire. The green, red, and blue dots represent the components G1, G2, and G3, respectively. The vertical dotted lines and the double-bar arrows between them mark the seven periods used to extract the acceleration parameters listed in Table~\ref{tab:fitres}. We note that when the fit was performed with two Gaussians, the most blueshifted component was assigned to G3.
\label{fig:fitres}}
\end{figure}

Figure~\ref{fig:spfit} presents the Gaussian fit results for the series of spectra just after the time sequence shown in Fig.~\ref{fig:acc_sp}. In panels~(a) to (h), the fitting describes the observed spectra with three components. Several impulsive peaks are identified, as shown in the \tv~plots in Fig.~\ref{fig:wtplot}. The positions of the spectral features marked by arrows in Fig.~\ref{fig:wtplot} correspond to 20:58:43~UT (panel~a), 20:59:12~UT (panel~d), 20:59:40~UT (panel~f), and 21:00:17~UT (panel~h). The fitting results for panels (a), (f), and (h) show the strong blueshifted features reaching velocities of approximately $-$170 to $-$200\,\kmps. In panel (d), a faint enhancement is present near $-$200\,\kmps\ (marked with a black arrow), suggesting that additional components may be required for a complete decomposition. In contrast, the spectra at panels~(b), (c), (e), and (g) located between the two peaks appear darker and exhibit a slightly smaller line width, indicating that the strong blueshifted component appeared only in one frame and disappeared in the subsequent frame.

Figure~\ref{fig:fitres} shows the results of the multi-Gaussian fitting during the time period from 20:49 UT to 21:29 UT. We performed the Gaussian fitting after 20:49~UT, when the jet became clearly distinguishable from the background. The fast burst ejections are well described by two or three Gaussian components, as described above and displayed in Figs.~\ref{fig:acc_sp} and \ref{fig:spfit}. The fitting results exhibit rapidly varying behavior as shown in Fig.~\ref{fig:spfit}, especially around 20:58--21:05~UT, which is the most dynamic phase of the jet. After this period, the variations become more or less monotonic.

During the interval from 21:10 to 21:14~UT, when the jet is comparatively stationary, the average velocities of the G2 and G3 components ($v_2$ and $v_3$) are around $-$26 and $-$107\,\kmps, respectively. The average line widths ($\sigma_2$ and $\sigma_3$) are 151 and 323\,m\AA, corresponding to nonthermal speeds of approximately 32 and 69\,\kmps, respectively. We note that the nonthermal speed of G3 of 69\,\kmps\ exceeds the normal nonthermal speed observed in the transition region \citep{depontieu_2015}.

\subsection{Acceleration of the jet's plasma}
Figure~\ref{fig:fitres} shows the acceleration periods of seven packets; these intervals were selected because the most blueshifted component showed a clear, nearly linear increase in Doppler velocity over time. Table~\ref{tab:fitres} summarizes the fitting results for G3 during each acceleration phase. The average acceleration for the seven listed cases is 0.9\,\kmpss. Assuming a vertically ejected jet (see Sect.~\ref{subsec:energy}), the acceleration is estimated to be around 1.3\,\kmpss. During the acceleration, the intensity amplitude ($A_3$) exhibits a systematic increase in all cases, whereas the Doppler width ($\sigma_3$) does not show any trend. 

\begin{table*}[ht!]
\caption{Linear slopes of the temporal evolution of the fitting parameters derived from the Si\,\textsc{iv} 1402.77\,\AA\ profiles for each time interval $(t_i, t_f)$. }
\label{tab:fitres}
\centering
\begin{tabular}{ccccc}
\hline\hline
$t_i$ & $t_f$ & $ dv_3/dt$ [km\,s$^{-2}$] & $d\sigma_3/dt$ [m\AA\,s$^{-1}$] & $dA_3/dt$ [DN\,s$^{-1}$]\\
\hline
20:57:19 UT & 20:58:25 UT & $1.3\pm0.2$ & $2.7\pm0.4$ & $2.9\pm0.5$ \\ 
21:00:54 UT & 21:01:51 UT & $1.2\pm1.6$ & $1.0\pm1.7$ & $3.8\pm3.1$ \\ 
21:05:54 UT & 21:06:50 UT & $1.1\pm0.3$ & $-2.5\pm1.0$ & $3.4\pm0.6$ \\ 
21:08:52 UT & 21:10:16 UT & $0.5\pm0.1$ & $1.3\pm0.3$ & $3.4\pm1.1$ \\ 
21:14:10 UT & 21:15:34 UT & $0.6\pm0.1$ & $-0.2\pm0.4$ & $3.1\pm1.2$ \\ 
21:19:56 UT & 21:21:30 UT & $0.5\pm0.1$ & $1.2\pm0.5$ & $0.5\pm0.7$ \\ 
21:26:11 UT & 21:27:26 UT & $1.2\pm0.1$ & $-2.7\pm0.6$ & $1.0\pm0.5$ \\ 
\hline
\end{tabular}
\tablefoot{The velocity slopes, line-width, and amplitude slopes of the blueshifted bursty component are listed. Uncertainties represent the $1\sigma$ standard errors of the linear fits.}
\end{table*}

The acceleration of the jet's plasma has been inferred previously from either apparent or Doppler velocities \citep{young_2014}. Our study provides the first direct observational determination of the jet acceleration from the temporal evolution of spectral line profiles, accounting for the 3D configuration derived from the Doppler-shift and apparent-motion estimates.

\subsection{Estimation of the kinetic energy release rate of blob ejections}\label{subsec:energy}

The emission of the O\,{\sc iv} 1399.77~\AA\ and O\,{\sc iv} 1401.16~\AA\ lines, which are commonly used to estimate the electron density in the transition region, are detected during the observation. However, the measured intensity ratio is near 0.5, which is beyond the density-sensitive range of the diagnostic \citep{zanna_2018,young_2018}, and the profiles of the two lines are irregular and inconsistent. This may be due to the complex velocity structures and line blending in the profiles.
Therefore, we adopted the known electron density range $10^{11}-10^{12}\,\mathrm{cm^{-3}}$ following several previous studies of similar jet structures \citep[e.g.,][]{chifor_2008,cai_2019,peter_2019,daniel_2021} for the following energy estimations.

We estimate the line-of-sight velocity of one of the plasma blob-like features expelled at 20:58~UT to be $\sim$$-$200\,\kmps~(shown as a dash-dotted blue line in Fig. \ref{fig:spfit}a). Its corresponding apparent motion in the SJ images is $\sim$230\,\kmps~(see the first slope in Fig.~\ref{fig:stplot}c2). Considering the projection angle ($\theta$) of 44$^\circ$ and the radially expelled jet (see the white arrow in Fig. \ref{fig:fov}), the blob is inferred to be ejected with a velocity of $\sim$310\,\kmps. Using the observed width of 0.76\arcsec\ and length of 3.3\arcsec\ of the blob-like feature, its volume is estimated to be $3.8\times 10^{23}$\, cm$^3$, assuming an ellipsoidal shape. Adopting an electron number density in the range of $10^{11}$--$10^{12}$\,cm$^{-3}$ and assuming a fully ionized hydrogen plasma, the total blob mass is estimated to be $6.4\times10^{10}$--$6.4\times10^{11}$\,g and the corresponding kinetic energy is the range of $3.1\times10^{25}$--$3.1\times10^{26}$\,erg.

The following blob is ejected after approximately $30$\,s. Using this recurrence time, we estimate a kinetic energy release rate of $\sim$1.0$\times$10$^{24}$--1.0$\times$10$^{25}$\,erg\,s$^{-1}$. It should be noted that the blob length (3.3\arcsec) is measured at a location approximately $6$\arcsec~above the slit position (see Fig. \ref{fig:blobsz}a) and, therefore, the kinetic energy and energy release rate are likely lower than these values. 

We emphasize here that the distance between the top of the dome and the slit position in the SJIs is approximately $1$\arcsec, corresponding to roughly 1000\,km when the viewing angle is taken into account. The distance 1000\,km is too short for the blob-like features to be accelerated with only 1.3\,\kmpss\ at the ejection location. Thus, the jet must have been ejected with a higher velocity at the reconnection site, localized region below the slit position. If the feature is linearly accelerated to $310$\,\kmps~over 1000\,km, the required acceleration is approximately $47$\,\kmpss, and the time required for the acceleration is $7$\,s. This implies that the jet acceleration is strong at the localized reconnection site and decreases away from it.

\section{Summary and discussion}
\label{sec:discussion}

The present study provides unique, detailed, and, most importantly, spectroscopic insight into jet dynamic evolution that is crucial for understanding the drivers and formation mechanism(s) of these phenomena. We report several fine morphological and kinematic details of this process using rare sit-and-stare spectroscopic data that provide particularly important information near a plasma density structure (plasmoid or blob-like-feature) release site, assumed to be a magnetic reconnection site, which is crucial for validating jet models.

We investigated the evolution of a dome-spire-shaped jet consisting of cool and hot components, accounting for the 3D configuration derived from the Doppler-shift and apparent-motion estimates. The hot jet identified in the UV/EUV imaging channels consists of blob-like features that are expelled along the jet spire. The blob-like features are ejected periodically at a cadence of $\sim30$\,s, with a speed of $\sim$310\,\kmps. The volume of the features is $3.8\times 10^{23}$\,cm$^3$. Based on these results, we estimate a kinetic energy release rate of $\sim$1.0$\times$10$^{24}$--1.0$\times$10$^{25}$\,erg\,s$^{-1}$, assuming an electron number density in the range of $10^{11}$--$10^{12}$\,cm$^{-3}$.
The cool component of the jet observed in \caii~line spectra shows a single velocity component. This suggests that the energy release at the site is caused by a single event rather than successive energy-release events. 

The most notable result of the present study is the observation of multiple periodic blob-like features with a cadence of $\sim$30\,s. This periodic blob-like occurrence is far shorter than previously reported (e.g.,
2--5\,min, \citealt{cheng_2018}; ${<}2$\,min, \citealt{joshi_2020}). We note that many previous studies do not clearly distinguish between jet recurrence and the ejection of individual blobs within a single jet event. Here, we emphasize that our observations also capture a series of blob-like ejections grouped into well-defined packets, rather than the recurrence of separate jets.

Based on our results, we infer a periodic expulsion of blob-like features and energy release at a localized region that can possibly be interpreted as a signature of the plasmoid-mediated magnetic reconnection in a dome-spire structured jet. The sudden emergence and disappearance of the fast blueshifted components in successive frames, as seen in Fig.~\ref{fig:spfit}, resemble the formation and escape of plasmoids predicted in numerical simulations \citep[e.g.,][]{bhattacharjee_2009, ni_2015}. These simulations show that an x-point null evolves into a Sweet-Parker-type current sheet, where the reconnection rate is enhanced by tearing-mode instability accompanied by plasmoid formation \citep{bhattacharjee_2009,edmondson_2010}. This process significantly enhances the reconnection rate beyond what classical models predict, thereby accounting for the observed impulsive energy release.

However, the interpretation of the observation of recurrent blob-like plasma ejection as the result of plasmoid-mediated reconnection should also be considered with caution. While 2D simulations describe plasmoids as closed magnetic islands, realistic 3D fan-spine configurations are expected to produce thread-like structures or mini-fluxropes \citep{yan_2022}. The ball-like appearance of these blobs in our observations may be the projection effect of such 3D filamentary structures. Furthermore, in contrast to plasma acceleration via magnetic tension force in 2D magnetic reconnection, realistic 3D magnetic reconnection allows more complex dynamics, such as continuous field-line-slipping reconnection without changing the field configurations at the separatrix or quasi-separatrix layers \citep{pontin_2022}. Further investigation is necessary to understand the underlying physical mechanisms of such phenomena.

Another possible explanation for the observed periodic phenomenon is recurring magnetic reconnection \citep[also suggested in][]{madjarska_2011} or oscillatory reconnection \citep{schiavo_2025}. Magnetic reconnection in the transition region, which is a steeply stratified density and temperature medium, may generate dynamically mixed and multi-temperature plasma flows \citep{li_2025}, because the height of the reconnection site can determine the plasma temperature \citep{shibata_1992,peter_2019}.

We note that the characteristic timescale of the blobs ($\sim30$\,s) is significantly longer than the elementary bursts reported in hard X-ray, microwave, and \halpha~wing emission \citep{qiu_2006,wang_2000,ding_2001}, which typically occur on millisecond timescales. This discrepancy suggests that the observed periodic, blob-like ejections represent a larger-scale manifestation of magnetic reconnection dynamics, rather than the fundamental microscopic energy release, which may manifest as elementary bursts. Future observations with higher temporal cadence and spectral resolution will be crucial for unraveling the detailed processes underlying elementary bursts and the observed periodic ejections.

Although magnetic reconnection is difficult to investigate directly because measuring the strength and topology of the magnetic field in the corona is currently not possible, indirect measurements such as Doppler shifts, nonthermal spectral line broadening, or multi-temperature emission evolution and morphology provide important constraints on the reconnection process. However, future instrumentation targeting high-cadence, multiwavelength spectroscopy is needed to help unravel finer details of the internal dynamics of magnetic reconnection on the Sun.

\begin{acknowledgements}
We thank the referee for the constructive and detailed comments, which have helped us substantially improve the manuscript. This work was supported by the National Research Foundation of Korea(NRF) grant funded by the Korea government (MSIT) (RS-2022-NR071796). M.M. acknowledges DFG grant WI 3211/8-2, project number 452856778. M.M. thanks the International Space Science Institute, Bern, Switzerland, for the Visiting Scientist grant.
K.-S. L. was supported by the National Research Foundation of Korea (RS-2025-23523356). IRIS is a NASA small explorer mission developed and operated by LMSAL with mission operations executed at NASA Ames Research center and major contributions to downlink communications funded by ESA and the Norwegian Space Centre. BBSO operation is supported by a US NSF AGS 2309939 grant and the New Jersey Institute of Technology. The GST operation is partly supported by the Korea Astronomy and Space Science Institute and the Seoul National University. SDO is a mission for NASA’s LWS program. AIA is an instrument on board the Solar Dynamics Observatory. All SDO data used in this work are available from the Joint Science Operations Center (http://jsoc.stanford.edu) without restriction.
\end{acknowledgements}

\bibliographystyle{aa} 
\bibliography{aa61497-26}

\begin{appendix}

\section{Jet-associated photospheric magnetic field configuration in the jet footpoints \label{appx:mag}}

\begin{figure}[ht!]
\includegraphics[width=\hsize]{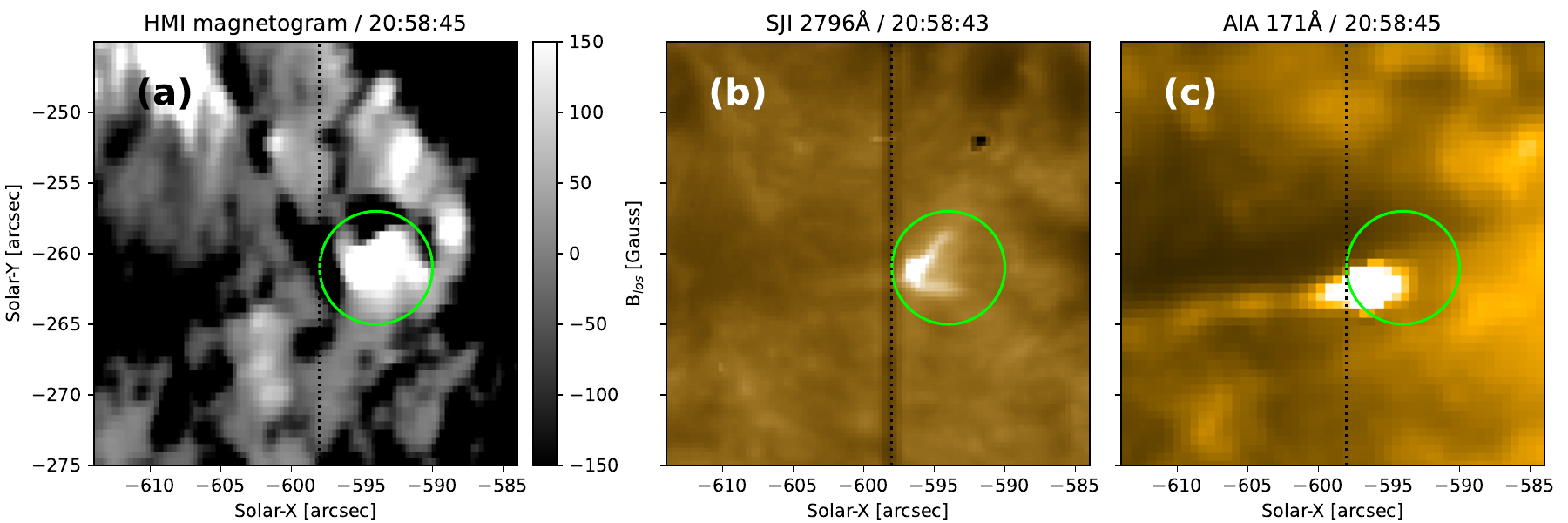}
\caption{Photospheric magnetic field (a) associated with the jet seen in the IRIS SJ 2796~\AA\ image (b), and the 171~\AA\ channel of SDO/AIA (c). Green circles indicate the location of the dome structure clearly seen in panel (b).
}
\label{fig:mag}
\end{figure}

\section{Background subtraction using Gaussian filtering \label{appx:gaussian_filter}}
\begin{figure}[ht!]
\includegraphics[width=\hsize]{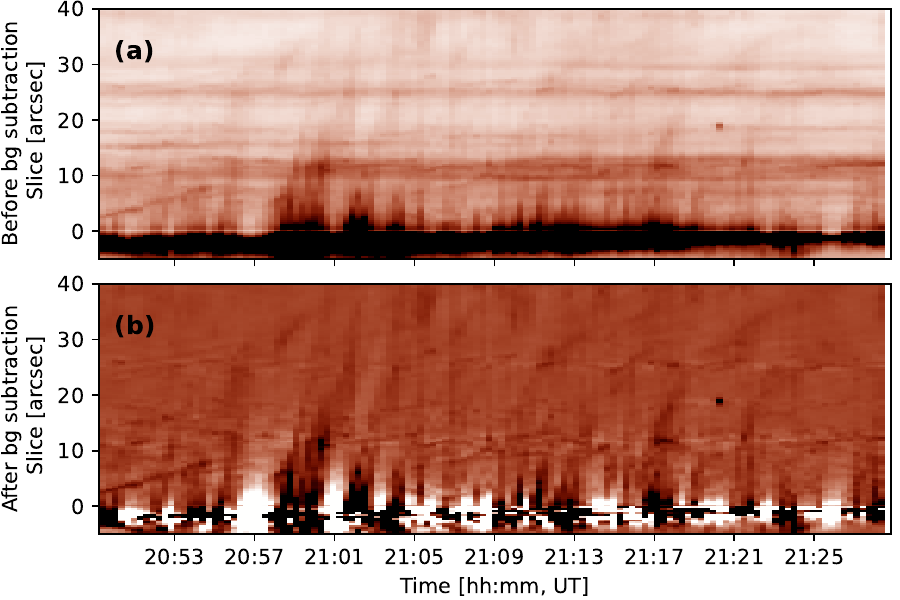}
\caption{Time-slice ($t$--$s$) plots of IRIS/SJI 1400\,\AA~images before (a) and after (b) background subtraction. We generated the background by applying a 1D Gaussian filter along the time axis with a standard deviation of $\sigma=5.6$~pixels (corresponding to 105\,s).}
\label{fig:gaussian_filter}
\end{figure}

\section{Size of the blob-like features\label{appx:blobsz}}

\begin{figure}[ht!]
\includegraphics[width=\hsize]{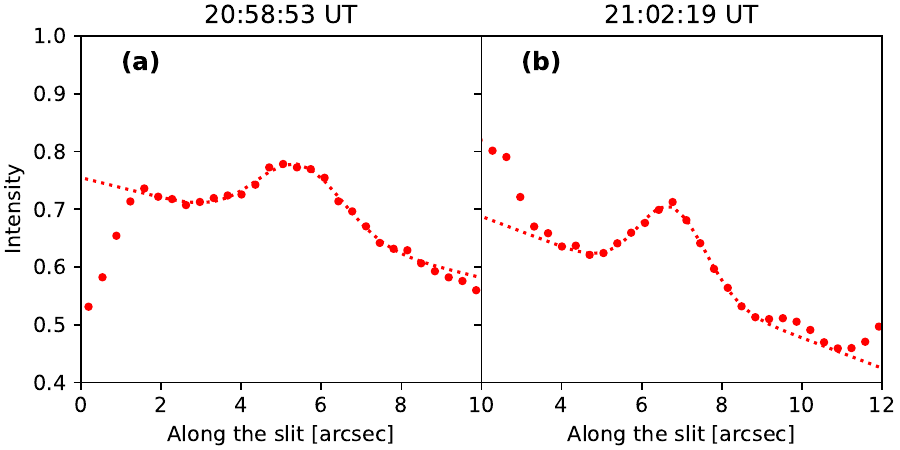}
\caption{Brightness distributions along the image slice at two time positions of the blob-like features shown in Fig.~\ref{fig:wtplot}, together with the corresponding Gaussian fitting results. The full widths at half maximum of the Gaussian fits are measured as 3.2\arcsec\ and 2.3\arcsec, respectively.
\label{fig:blobsz}}
\end{figure}

We measured the length of two blob-like features by fitting the image slice intensity to a Gaussian, as shown in Fig. \ref{fig:blobsz}. The full width at half maximum of the Gaussian fits is measured as 3.2\arcsec\ and 2.3\arcsec, respectively. This is larger than the width of the spire of $\sim$0.76\arcsec (see Sect. \ref{morphology}), meaning that the blobs have an elongated structure. This is consistent with \citet{zhang_2016}, who reported that some observed blob-like features exhibit elliptical shapes. We also note that the measured size is within the range of plasma blob-like features reported in various EUV and coronal jets \citep{chen_2022,zhang_2019,mulay_2023}.

\end{appendix}
\end{document}